\documentclass[%
 reprint,
 superscriptaddress,
 amsmath,amssymb,
 aps,
 multiline,
 prl,
]{revtex4-2}

\usepackage{graphicx}% Include figure files
\usepackage{dcolumn}% Align table columns on decimal point
\usepackage{bm}% bold math
\usepackage{svg}
\usepackage{titlesec}
\renewcommand{\selectlanguage}[1]{}

\definecolor{darkblue}{rgb}{0,0,0.6}
\definecolor{darkred}{rgb}{0.6,0,0}
\usepackage[colorlinks=true,urlcolor=darkblue,citecolor=darkblue,linkcolor=darkred]{hyperref}

\DeclareMathOperator{\sech}{sech}

\begin{document}

%\title{Green's function neural networks: How to probe non-linear physics in experiments}
%\title{Neural operators learn how non-reciprocity creates and regularizes shocks from data} %in non-linear systems}

%
%\title{Neural operators learn how non-reciprocity creates shocks from data} 
%DB
%\title{Neural operators learn  how non-reciprocal interactions propagate and stabilize solitary waves} 
%VV
%\title{Neural operators learn  how non-reciprocal interactions give rise to solitary waves} 
%VV
%\title{Interpreting neural operators helps to discover mathematical models that govern  nonlinear dynamics experiments}
%\title{How to interpret neural operators to discover mathematical models governing  nonlinear dynamics experiments}
%DB
%\title{Interpreting neural operators to model nonlinear-physics experiments
%: how hydrodynamic interactions propagate solitary waves}
%\title{Interpreting neural operators: how non-reciprocal interactions propel nonlinear waves}
\title{Interpreting neural operators: how nonlinear waves propagate in non-reciprocal solids}

%in non-linear systems}
%\title{Neural operators reveal how non-reciprocity creates and regularizes shocks} %in non-linear systems}
%\title{Learning to regularize experimental singularities using Green's function neural networks}
%\title{Green's function neural networks reveal how non-reciprocity regularizes nonlinear shock waves in 1D crystals}
%\title{Green's function neural networks reveal how non-reciprocity regularizes  shock singularities}
%\title{Green's function neural networks reveal how microscopic non-reciprocity smooths out shock waves}
%\title{Learning how Microscopic Non-Reciprocity  smothe Shock Wave Smoothing: Insights from Green's Function Neural Networks}

\date{\today}

\titleformat{\subsection}[runin]{\normalfont\itshape}{}{0pt}{}[ ---]
\author{Jonathan Colen}%$^{*}$}%
\thanks{These two authors contributed equally}
\affiliation{James Franck Institute, University of Chicago, Chicago, IL 60637}
\affiliation{Department of Physics, University of Chicago, Chicago, IL 60637}
\affiliation{Joint Institute on Advanced Computing for Environmental Studies, Old Dominion University, Norfolk, Virginia 23539}
\author{Alexis Poncet}%$^{*}$}%
\thanks{These two authors contributed equally}
\affiliation{ Univ. Lyon, ENS de Lyon, Univ. Claude Bernard, CNRS, Laboratoire de Physique, F-69342, Lyon.}
\author{Denis Bartolo}%
\email{denis.bartolo@ens-lyon.fr}
\affiliation{ Univ. Lyon, ENS de Lyon, Univ. Claude Bernard, CNRS, Laboratoire de Physique, F-69342, Lyon.}
\author{Vincenzo Vitelli}
\email{vitelli@uchicago.edu}
\affiliation{James Franck Institute, University of Chicago, Chicago, IL 60637}
\affiliation{Department of Physics, University of Chicago, Chicago, IL 60637}

\begin{abstract}
We present a data-driven pipeline for model building that combines interpretable machine learning, hydrodynamic theories, and microscopic models. 
The goal is to uncover the underlying processes governing nonlinear dynamics experiments.
We exemplify our method with data from microfluidic experiments where crystals of streaming droplets support the propagation of nonlinear waves absent in passive crystals. 
By combining physics-inspired neural networks, known as neural operators, with symbolic regression tools, we generate the solution, as well as the mathematical form, of a nonlinear dynamical system that accurately models the experimental data. 
Finally, we interpret this continuum model from fundamental physics principles. 
Informed by machine learning, we coarse grain a microscopic model of interacting droplets and discover that  non-reciprocal hydrodynamic interactions stabilise and promote nonlinear wave propagation.% whose propagation is otherwise hindered by contact interactions. 

\end{abstract}

\maketitle

%Big focus: framing the problem is IMPORTANT
%How do we structure a neural network about a physical hypothesis? What are the ingredients of a problem that can be approached with this method

%Add in difference between physics-informed neural networks and physics-inspired neural networks

\subsection{Introduction.}
The goal of physical modeling is to build a set of rules which can predict and explain behaviors observed in experiments. 
In continuum theories, these rules are typically grounded in symmetries and conservation laws, which determine the relevant degrees of freedom and the equations they must satisfy.
%Unfortunately, in out-of-equilibrium systems, the number of parameters may be too large.
%For example, they lead to the dynamical equations for velocity in fluid mechanics or the equations governing displacements in linear elasticity. 
This approach faces challenges when applied to experiments that only reveal the dynamics of a finite subsets of all the interacting degrees of freedom. 
%whose behavior is driven by complex interactions at the microscopic scale.
This situation is not exceptional, in particular it is the norm when soft matter is driven out-of-equilibrium \cite{Ramaswamy2001,marchetti_hydrodynamics_2013}.
In these systems, the particles, such as colloids, droplets, or bubbles continuously exchange energy and momentum with a solvent. 
As a result they exhibit a variety of behaviors, driven by effective interactions that
 violate microscopic reversibility constraints such as detailed balance \cite{fruchart_2023,scheibner_odd_2020} or Newton's third law ~\cite{ivlev_statistical_2015,Poncet2022,brauns_non-reciprocal_2023,saha_effervescent_2022,saha_scalar_2020,you_nonreciprocity_2020,mandal_robustness_2022,avni_non-reciprocal_2023,weis_exceptional_2023,martin_transition_2024,fruchart_non-reciprocal_2021}.
Such complications require new approaches to model building that can identify and predict the effects of these microscopic interactions in experiments \cite{li_data-driven_2019,VanSaarlos2024,seara_sociohydrodynamics_2024,supekar_learning_2023}.

%The continuum mechanics approach faces challenges when applied to living systems, where complex behavior is influenced by proteins at the microscopic scale. 
%The significant role of such structures, which exist outside the framework of symmetries and conservation laws, make it difficult to both find suitable variables and to identify rules and interactions governing those variables.
%{\color{red} (Complexity of \emph{out-of-equilibrium systems}, \emph{non-reciprocal interactions}, and examples closely related to our system.)}

Data-driven methods, which have shown much promise in learning physical models from observation data, have the potential to overcome these challenges.
Deep neural networks~\cite{lecun_deep_2015,schmidhuber_deep_2015} can learn to accurately predict behavior for a variety of physical~\cite{raissi_physics-informed_2019,carleo_machine_2019,cichos_machine_2020,karniadakis_physics-informed_2021,carrasquilla_machine_2017,bapst_unveiling_2020,brunton2020machine,zhou_machine_2021,colen_machine_2021} and biological~\cite{jumper_highly_2021,soelistyo_learning_2022,zaritsky_interpretable_2021,schmitt_machine_2024,lefebvre_learning_2023} systems.
More recently, neural operators have been introduced to directly learn the solution operator for partial differential equations~\cite{li_fourier_2021}.
When incorporated within deep neural networks, they are capable of forecasting the behavior of complex dynamical systems such as the weather ~\cite{li_fourier_2021,pathak2022fourcastnet}.
While these methods are invaluable for efficiently and accurately predicting dynamics from data, less is known about how to interpret what they have learned about a physical system. 

In this case study, we demonstrate how interpreting neural operators can lead to data-driven model discovery by focusing on a paradigmatic out-of-equilibrium hydrodynamic problem: particles driven in a surrounding fluid~\cite{Beatus2012}. 
%Beatus, T., Bar-Ziv, R. H. & Tlusty, T. The physics of 2D microfluidic droplet ensembles. Phys. Rep. 516, 103–145 (2012).
Such systems exhibit a variety of phenomena: phonons in overdamped systems~\cite{Beatus2007,Desreumaux_2013}, shock formation~\cite{Beatus2009,champagne_stability_2011},
%Beatus, T., Tlusty, T. & Bar-Ziv, R. Burgers shock waves and sound in a 2D microfluidic droplets ensemble. Phys. Rev. Lett. 103, 114502 (2009).
and symmetry-dependent melting of crystal phases~\cite{saeed_2023}.
We build on an experimental set-up in which water droplets %(140~$\mu$m of diameter),
 advected inside a microfluidic channel  form a propagating shock due to hydrodynamic interactions.
%Hydrodynamic interactions between the droplets, whose speed is smaller than that of the surrounding fluid, induce and propagate shock waves.
Using a neural network with an internal neural operator layer~\cite{li_fourier_2021}, we transform the experimental movie into a set of variables whose dynamics are solved by an unknown partial differential equation. 
%By applying an encoder/decoder neural network to the experimental movies, we learn a set of features governed by an unknown PDE while a Fourier neural operator learns the solution to that PDE. 
Using sparse regression~\cite{brunton_discovering_2016,kaheman_sindy-pi_2020,champion_unified_2020}, we then discover the equation that these machine-learned variables satisfy. 
We find that the discovered continuum model contains an essential term that had remained overlooked.  
Informed by our data-driven method, we then trace this term to the non-reciprocal nature of hydrodynamic interactions by solving suitable microscopic models.
Beyond the specifics of our microfluidic experiments, we  show how microscopic non-reciprocity generically results in the propagation of macroscopic solitons through non-equilibrium crystals.
Our approach demonstrates how combining interpretable neural operators with theoretical modeling can provide previously-unknown insights into the physics governing complex non-linear systems.

\begin{figure*}
    
    \centering
    \includegraphics[width=0.8\textwidth]{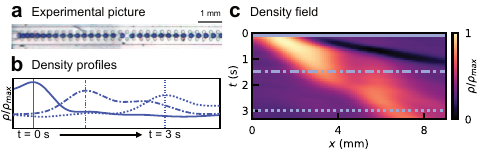}
    \caption{{\bf Coarse-graining microfluidic particle experiments.}
        %Suggction: two column figure. Top left a. Bottom left b. light grey-->same blue as the droplets. darker lines--> dotted and dashed lines. Right c. Remove d. % JC: Done
        (\textbf{a}) Experimental picture of a microfluidic particle stream~\cite{champagne_stability_2011}.% which is coarse-grained into a local density field. 
        (\textbf{b}) Experimental pictures are coarse-grained into a density profile and normalized by the maximum density $\rho_{max}$. 
        %Dynamics of coarse-grained density in response to an initial jam at $t = 0$. 
        (\textbf{c}) Spatio-temporal evolution of the density field shows formation of a shock front in response to an initial jam.
        %(\textbf{d}) Hydrodynamic interactions of droplets in a confined channel lead to non-reciprocal interactions.
        %\textcolor{red}{Suggestions. Remove (d) since it seems that we don't need it. Use a color similar to the droplets in (a) (blue) for the cuts in (b) and (c).} %JC: Done
        %\textcolor{red}{
        %    (a) Add scale bar, change movie to picture % Done
        %    (b) Add label "Density profile", add units to $\rho$ and $t$ %Done
        %    (c) Add label "Density field", make lines a lighter blue %Done
        %}
        }
    \label{fig:experiments}
\end{figure*}

\subsection{Learning experimental dynamics.}
% Hypothesis testing and model identification for a nonlinear system with unknown physics
To learn macroscopic physics from microscopic data,  
we consider an experimental movie~\cite{champagne_stability_2011} in which a one dimensional stream of  water droplets (diameter $140\,\rm \mu m$) are advected by an organic solvent inside a microfluidic channel, see~\cite{champagne_stability_2011} for details. 
The droplet stream
develops a shock in response to a jam induced at $t = 0$ (Fig.~\ref{fig:experiments}a).

Coarse-graining the droplet positions yields a density field $\rho(x,t)$ which shows the shock deforming and propagating through space and time (Fig.~\ref{fig:experiments}).
The system behavior is driven by microscopic interactions between the droplets, which may include a simple nearest-neighbor repulsive interaction as well as hydrodynamic backflows.
Previous studies modeled this experiment at the continuum level to leading order in derivatives and non-linearities. 
The resulting model reduced to a Burgers' equation obeyed by the droplet density $\rho(x,t)$~\cite{champagne_stability_2011}
\begin{align}
    \partial_t \rho + (c - \alpha\rho) \partial_x \rho = 0,
    \label{eq:densityburgers}
\end{align}
where $c$ is a basic advection speed, and $\alpha$ is a non-linearity that creates shocks.
Here, without making any assumption about the relevant interactions, we learn to predict the dynamics of this movie using a deep neural network. 
To train the network, we segmented the experimental movie into overlapping $3\,$s$\,\times \,9\,$mm training examples. For each example, the network receives the one-dimensional slice $\rho(x) = \rho_0 (x)$ and predicts the full $\rho(x, t)$ response to that initial condition (Fig.~\ref{fig:neuraloperator}a). 
The network architecture has three components: (i) an input convolutional block which transforms the experimental density field $\rho_0(x)$ into a latent variable $\phi_0(x)$, (ii) a single-layer Fourier neural operator which evolves $\phi$ in time according to an unknown \textit{linear} partial differential equation, and (iii) an output convolutional block which transforms the predicted $\phi(x, t)$ field into a predicted density field $\rho(x, t)$. 
Both the input and output blocks used 1-D convolutional layers which only use spatial information to construct features. 
The full architecture is shown in Fig.~\ref{fig:neuraloperator}a.
We note that the dynamics were solely  predicted by the central neural operator layer. However, while the neural operator itself is linear, the input and output transformations enable the network to predict nonlinear dynamics~\cite{gin_deepgreen_2021}. 
Once trained, the neural network is capable of predicting experimental dynamics including the propagation of the initial jam in space and time from past to present (Fig.~\ref{fig:neuraloperator}a-c).

\subsection{Interpreting the neural operator.}
\begin{figure*}
    \centering
    \includegraphics[width=\textwidth]{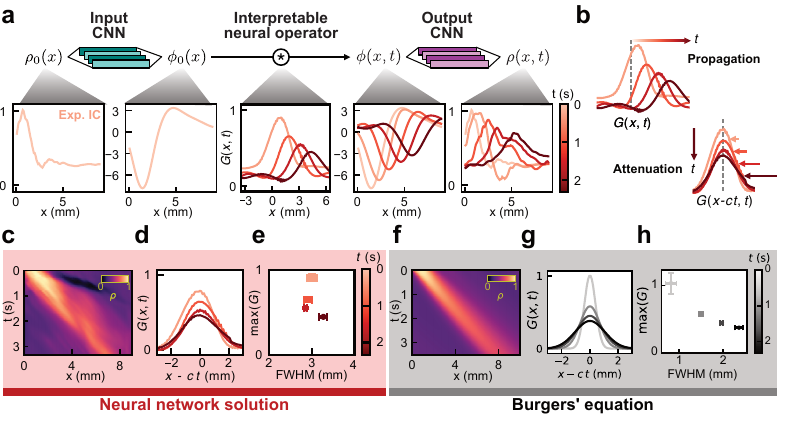}
    \caption{Learning the dynamics of shocks in microfluidic particles experiments.
    (\textbf{a}) A neural network forecasts the particle density field $\rho(x)$ from initial conditions. The network uses a 1-D convolutional block to map the density field to an internal variable $\phi(x)$. A single-layer linear Fourier neural operator forecasts the evolution of $\phi$. A second 1-D convolutional block converts $\phi(x, t)$ into a predicted density field $\rho(x, t)$.
    (\textbf{b}) The neural operator is interpretable as a Green's function for an unknown linear PDE which defines the response to an perturbation. The learned Green's function predicts a response which propagates in space and attenuates over time.
    (\textbf{c}) Neural network predictions for the density field $\rho(x, t)$. %The error rate is reported in Table~\ref{tab:error_rates}.
    (\textbf{d}) Plot of the machine-learned Green's function in the co-moving frame. 
    (\textbf{e}) Comparing the magnitude and width of the kernel $G(x, t)$ shows that the response to a perturbation will decrease in magnitude over time, but not disperse.
    (\textbf{f-h}) Exact solution to Burgers' equation from experimental initial conditions. $G(x,t)$ is a diffusion kernel applied after a Cole-Hopf transformation of the density $\rho$. The neural network learns distinct physics from Burgers', which exhibits spreading due to viscous dissipation.
    }
    \label{fig:neuraloperator}
\end{figure*}

The single-layer linear Fourier neural operator can be interpreted as a Green's function for the internal variable $\phi$. 
This learned Green's function predicts how the response to an initial perturbation will propagate in space over time (Fig.~\ref{fig:neuraloperator}a-b). 
In the co-moving frame, defined by the spatial coordinate $x - c t$ where $c$ is the speed of the wave, we observe that the response decreases in magnitude while the width stays nearly constant (Fig.~\ref{fig:neuraloperator}c-e). 
The encoder/decoder architecture transforms the nonlinear dynamics of the density $\rho$ into a linear problem that can be solved using this Green's function~\cite{gin_deepgreen_2021}. 

A similar approach can exactly solve the Burgers' equation \eqref{eq:densityburgers} previously proposed for this system. 
Upon adding a small diffusion term $D\,\partial_x^2 \rho$ and effecting a Cole-Hopf transformation, we obtain a linear partial differential equation whose Green's function is that of the standard diffusion equation (see SI). In Figure~\ref{fig:neuraloperator}f-h, we plot the dynamics predicted by Burgers' equation from experimental initial conditions.
As a benchmark for our machine-learned solution, we trained an identical neural network on a synthetic movie of an exact solution to Burgers' equation (SI Fig.~S3).
While this model learned an internal variable transformation that was distinct from Cole-Hopf, the Green's function is qualitatively similar to the diffusion kernel with a response whose amplitude decays and width grows over time.

The neural network (Figure~\ref{fig:neuraloperator}c-e) more accurately predicts the experimental movie than an exact solution to Burgers' equation (Figure~\ref{fig:neuraloperator}f-h) and uses a qualitatively distinct Green's function to do so.
We stress that while both Green's functions decay in amplitude over time, the diffusion kernel has a response width that grows over time while the neural operator solution remains constant (Fig.~\ref{fig:neuraloperator}e,h). 
As a result,  the neural network shock remains sharp unlike the Burgers' shock that disperses  (Fig.~\ref{fig:neuraloperator}a,c,f).
The distinct behavior of the neural operator trained on experiments from both the exact and benchmark solutions to Burgers' equation suggests that the non-linear waves seen in experiments obey physics beyond Burgers' equation.

\begin{figure}[ht!]
    \centering
    \includegraphics[width=0.45\textwidth]{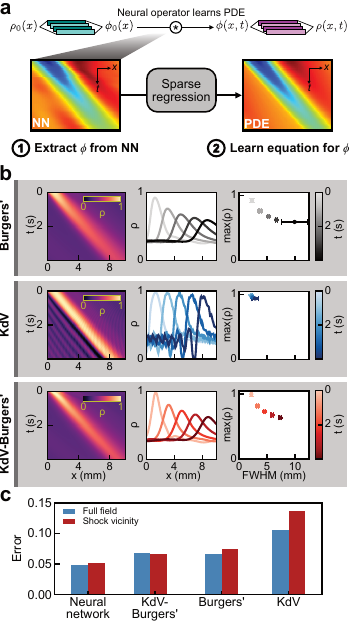}
    \caption{Interpreting the neural operator.
    (\textbf{a}) The neural operator evolves the internal variable $\phi(x, t)$ according to a partial differential equation. Using SINDy, we learn the PDE governing $\phi(x, t)$. The resulting equation (\ref{eq:burgersinternal}) includes a novel $\partial_x^3$ KdV term~\cite{champagne_stability_2011}.
    (\textbf{b}) Simulations of Burgers' equation (first row), KdV equation (second row), and KdV-Burgers' equation (third row) from experimental initial conditions. 
    %These equations are motivated by our interpretable ML pipeline. 
    For each equation, we plot the density field (left) and traces at fixed time points (middle). 
    %The KdV prediction separates into a superposition of soliton solutions.
    To characterize the shock front, we plot the height vs. width at different time points (right). 
    KdV-Burgers' preserves the sharpness of the shock and shows less spreading over time compared to Burgers'.
    (\textbf{c}) Error rates for neural network (Fig.~\ref{fig:neuraloperator}) and analytic solutions (Fig.~\ref{fig:sindy}). %Baseline is a constant uniform density field. 
    Blue bars are the mean absolute error $\langle |\rho - \rho_0 | \rangle$ for the full field, while red is the error within 1 mm of the shock front. Numerical values in SI Table S2.
    %\textcolor{red}{(Include table I as a bar plot: Fig 3c ?)}
    %\textcolor{red}{
    %    (c) Change error rate to "Mean Absolute Error" or just "Error" %Done
    %    (c) Move baseline to caption and refer to its numerical value in the SI %Done
    %}
    }
    \label{fig:sindy}
\end{figure}

To learn an equation governing the internal variable $\phi(x, t)$ (Fig.~\ref{fig:sindy}a), we resort to the SINDy method ~\cite{brunton_discovering_2016}, which has proven to effectively learn non-linear dynamical equations from exeprimental data~\cite{brunton2020machine,supekar_learning_2023} . 
As the linear neural operator ensures $\phi$ obeys a linear PDE, we used a library of $\phi$ and its derivatives and found the following equation
\begin{align}
    \partial_t \phi - 1.86\, \partial_x \phi = -0.31 \, \phi + 0.003\, \partial_x^2 \phi - 0.008 \, \partial_x^3 \phi
    \label{eq:burgersinternal}
\end{align}
Simulating this equation for $\phi(x, t)$ yields excellent correspondence with the neural network solution (Fig.~\ref{fig:sindy}a, $R^2=0.96$), indicating that it is indeed an excellent model for the dynamics solved by the neural operator.
We applied the neural network's output CNN to this predicted $\phi$ field to achieve an accurate prediction of the density dynamics (SI Fig.~S2).

The learned equation (\ref{eq:burgersinternal}) resembles a linearized KdV-Burgers' equation, containing a dispersive $\partial_x^3 \phi$ term characteristic of the Korteweg de-Vries equation. 
The $\partial_x \phi$ term captures both the basic advection term and a linearization of the $\rho \partial_x \rho$ term in \eqref{eq:densityburgers}. 
We do note that the model has learned a damping term whose importance to the learned dynamics is weaker than the dispersive $\partial_x^3 \phi$ term but stronger than the diffusive $\partial_x^2 \phi$ term (SI Table~S1).
The new KdV term in the dynamics of $\phi$, as well as the qualitatively distinct Green's function learned by the neural operator, corroborates our conclusion that the dynamics of the experimental density field is governed by  physics beyond Burgers' equation. 
Hence, we consider the more general nonlinear KdV-Burgers' equation:
% Put KdVB equation here
\begin{equation}
    \partial_t \rho + (c - \alpha \, \rho) \partial_x \rho = D \partial_x^2 \rho + \beta \, \partial_x^3 \rho %= 0
    \label{eq:kdvb}
\end{equation}
We used the experimental movie to directly fit coefficients to a Burgers' equation ($\beta=0$), a KdV equation ($D=0$), and a KdV-Burgers' equation \eqref{eq:kdvb}.
We plot the predictions and compare accuracy in Figure~\ref{fig:sindy}b-c. 
Both Burgers' and KdV-Burgers' achieve similar error rates on the entire field, although neither matches the performance of the neural network. 
The difference between the two is more apparent near the shock front. The KdV-Burgers' solution maintains a more pronounced shock over time (Fig.~\ref{fig:sindy}b) and achieves a lower error rate than Burgers' in the vicinity of this shock front (Fig.~\ref{fig:sindy}c).
%\textcolor{red}{(We should write KdVB eq somewhere.)}

\iffalse
\begin{table}[]
    \centering
    \begin{tabular}{c | c |c}
         \textbf{Solution} &  \textbf{MAE} & \textbf{Shock-MAE} \\
         \hline \hline
         Neural network &  0.049 & 0.051 \\
         \hline
         Burgers' & 0.067 & 0.074 \\
         KdV & 0.11 & 0.14 \\
         KdV-Burgers\' & 0.068 & 0.067 \\
         \hline 
         Baseline & 0.17 & 0.34
    \end{tabular}
    %No acronym
    \caption{ Error rates for neural network (Fig.~\ref{fig:neuraloperator}) and analytic solutions (Fig.~\ref{fig:sindy}). We compute the mean absolute error (MAE) $\langle |\rho - \rho_0| \rangle$ for the whole field and within a 1-mm region centered at the shock (Shock-MAE). Baseline reports the error rates of predicting a uniform density field given by the mean of the movie.
    }
    \label{tab:error_rates}
\end{table}
\fi
\begin{figure*}
    \centering
    \includegraphics[width=\textwidth]{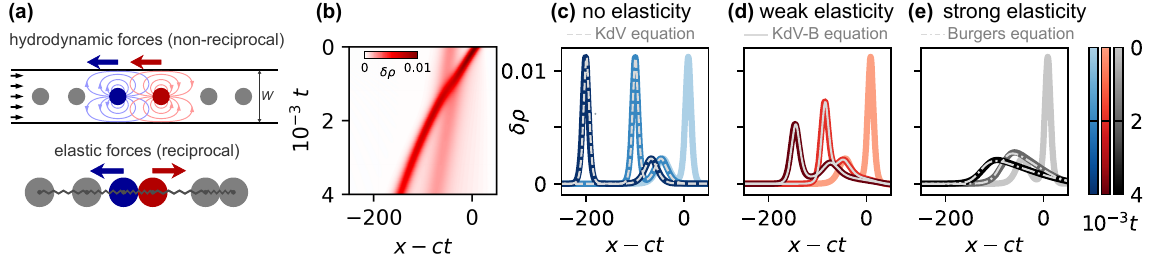}
    \caption{
    Microscopic modeling of the experiments and confirmation of the KdV-Burger behavior.
    (a) We model the droplets as an overdamped chain of beads that interact by two kinds of interactions: hydrodynamic forces which are non-reciprocal ($f^H_{m\to n} = f^H_{n\to m}$) and elastic forces which correspond to usual springs ($f^E_{m\to n} = -f^E_{n\to m}$). %nearest neighbor interactions:
    %hydrodynamic and elastic interactions. Hydrodynamic forces between two neighbors are symmetric ($f^H_{m\to n} = f^H_{n\to m}$) while elastic forces correspond to usual springs ($f^E_{m\to n} = -f^E_{m\to n}$).
    (b) We run numerical simulations starting from small density fluctuations $\delta\rho = \rho - a^{-1}$ corresponding to a two-soliton solution of the KdV equation. At weak elasticity ($k=0.02$) the solitons propagate by crossing one-another, and are damped by the elasticity.
    (c-e) The colored lines show the time-evolution %of the two-soliton solution
    in the simulations at no elasticity ($k=0$), weak elasticity ($k=0.02$, corresponding to panel b) and strong elasticity ($k=0.2$). In the three cases, this evolution is in excellent agreement with the numerical integration of the KdV-Burgers equation (shown in panel (d) with thin gray lines).
    Dashed lines in panels (c) and (e) are respectively the analytical solutions of the KdV equation and the Burgers equation, showing the two limit cases.
    %In panel (c), the dashed black line are the analytical solution of the two-soliton solution of the KdV equation. In panel (e), the dashed black line are the solution of the Burger's equation, showing that this equation reproduces well the strongly elastic regime.
    }
    \label{fig:microscopics}
\end{figure*}
\subsection{Interpretation of learned non-linear  dynamics from microscopics.}
The neural operator discovered a KdV-Burgers dynamics but provided no first-principles interpretation for it.
We now seek to understand its physical origin.
To make progress, we build a minimal model where the dynamics of the droplets is determined by the competition between  contact  and hydrodynamic interactions.  
The equations of motion for the droplet positions $R_n$ are
\begin{equation} \label{eq:micro_eq}
    \zeta(\partial_t R_n-v_0) = \sum_{m\neq n}\left( f_{m\to n}^H + f_{m\to n}^E \right), %f_H(R_n, R_m) +f_E(R_n, R_m)
    %k(R_{n+1}-2R_n+R_{n-1}),
\end{equation}
where $\zeta$ a friction coefficient and $v_0$ the advection speed of an isolated droplet. 
$f^E$ is an elastic force that models the repulsion between the soft droplets which we approximate by linear springs $f_{n+1\to n}^E = k(R_{n+1}-R_n-a)$ ($a$ is the lattice spacing and $k$ the elastic stiffness). 
$f^H$ models the hydrodynamic force experienced by a droplet in response to the backflows induced by the motion of its neighbors, see Fig. \ref{fig:microscopics}a. 
Unlike $f^E$,   hydrodynamic forces do not obey Newton's third law~\cite{Poncet2022}. They are non-reciprocal:  $f^H_{m\to n} = f^H_{m \to n} = f_H(R_n - R_m)$. 
In practice, we use the standard form
 $f_H(r) = A \,\mathrm{csch}^2 (\pi r/W)$, where $W$ is the channel width \cite{Beatus2007}.

%We assume that the dynamics of the droplets is overdamped and that they interact via two distinct type of forces:  repulsive interactions ($f^E$) that model the contact forces between the soft droplets, and hydrodynamic interactions 
%non-reciprocal hydrodynamic forces $f^H$, and weak repulsive interactions $f^E$
%that we model as springs of constant $k$ between nearest neighbors
%(Fig. \ref{fig:microscopics}a). The equations of motion for the droplet positions $R_n$ are

%with $\zeta$ the friction coefficient and $\tau$ the microscopic time. 
%The elastic term $f^E$ is a reciprocal interaction with $f^E_{m\to n} = - f^E_{n\to m}$  which for simplicity we consider to correspond to nearest neighbors springs with a lattice spacing $a$: $f_{n+1\to n}^E = -k(R_{n+1}-R_n-a)$.
%The hydrodynamic term $f^H$ strongly violates Newton's third law with $f^H_{m\to n} = f^H_{m \to n} = f_H(R_n - R_m)$, and we choose $f_H(r) = A q^2 \mathrm{csch}^2 (qr)$ (see SI for a discussion of other types of interactions).
%We choose $f_H(x) = A q_0^2 \mathrm{csch}^2(q_0x)$.
%This corresponds to interactions between driven intruders that behave as $A/r$ at short distance and are screened over a length $W=\pi/q$ corresponding to the width of the channel in the transverse direction \cite{Beatus2007}.
%Importantly, these effective interactions %are strongly non-reciprocal
%strongly violate Newton's third law: $f_H(-x)=f_H(x)$.
We perform numerical simulations of Eq.~\ref{eq:micro_eq} with periodic boundary conditions, $\zeta=a=1$, $W=2$, and $A=-(\pi/W)^2$.
To probe the relevance of KdV-Burgers physics, we choose initial conditions that correspond to a 2-soliton solution of the KdV equation (Fig.~\ref{fig:microscopics}b-e, see also SI for different initial conditions).
When ignoring the soft contact interactions ($k=0$), the particle dynamics is perfectly captured by a continuum model having the form of a KdV equation. Two solitions cross each other and propagate freely
(Fig. \ref{fig:microscopics}c). 
In the limit of strong elastic forces ($k=0.2$), the initial shape is not preserved, shocks develop and their dynamics is well described by a Burgers equation (Fig. \ref{fig:microscopics}e).
In the intermediate regime, where both contact and hydrodynamic forces compete on equal footing ($k=0.02$), we find that the propagation of the initial density fluctuation is quantitatively predicted by the  KdV-Burgers equation Eq.~\eqref{eq:kdvb}, (Fig. \ref{fig:microscopics}d).

We are now equipped to interpret the KdV-Burgers dynamics  revealed  by our neural operator.  
We focus on large distances ($\tilde x=\epsilon (n - ct)$ with $\epsilon\to 0$) and large times ($\tilde t = \epsilon^3 t$), we then can recast  Eq.~\eqref{eq:micro_eq} into a PDE obeyed by the density field $\rho(x,t)$. 
Informed by our Neural operator, we do not restrain our expansion to leading order in gradients and find  that the dynamics of $\rho(x,t)$  is ruled by a KdV-Burgers equation~\eqref{eq:kdvb}. 
We provide a detailed derivation of our continuum model in SI and generalize it beyond the specifics of hydrodynamic interactions.

This procedure makes it possible to understand the physical origin of all  terms in Eq.~\eqref{eq:kdvb}.
The damping ($D$) is solely controlled by the elastic interactions $f^E$.
Conversely,  the advection, the non-linear and the dispersive terms ($c$, $\alpha$ and $\beta$) originate from the hydrodynamic forces ($c =v_0 -2\sum_{m=1}^\infty m f_H'(ma)$, 
$\alpha = -2\sum_{m=1}^\infty m^2 f_H''(ma)$, and $\beta = \sum_{m=1}^\infty m f_H'(ma)/3$).
Unlike in ~\cite{Veenstra2024},  the dispersive and non-linear terms rooted in hydrodynamic interactions conspire to stabilize and propel solitary waves without relying on inertia.
%Few words to explain that hydro --> solitons via competition between shock and dispersion while contact results in damping.  

Our first principles modeling and machine learning algorithms play a mutually beneficial role. 
While the neural operator can suggest the presence of a KdV term, it could not provide a mechanism justifying its existence. 
Our simulations and theoretical argument help rationalizing these findings which had been hitherto overlooked. 
They provide insight on the origin of the KdV-like physics:
beyond the specifics of our driven emulsions, non-reciprocal interactions generically  stabilize and propagate solitary waves in overdamped one-dimensional systems.
%the higher-order dispersive term arises from non-reciprocal microscopic interactions between the droplets.
Neural networks' role as universal approximators provides an alternative route to characterizing experiments. In learning to predict dynamics, they build a ``maximal model'' which
%which one can interpret as an upper bound on what one can hope to capture.
%One can then use this maximal model and any interpretations it 
can then serve as a target for more principled theoretical investigations.
\begin{acknowledgments}
    This project received funding from the European Research Council (ERC) under the European Union’s Horizon 2020 research and innovation programme (grant agreement no. 101019141) (D.B. and A.P.).
\end{acknowledgments}

\bibliographystyle{apsrev4-1}
\bibliography{referencesAlexis.bib}

\end{document}

\title{Interpreting neural operators: how nonlinear waves propagate in non-reciprocal solids\texorpdfstring{\\[0.2cm]}{--} Supplementary Information}

\author{Jonathan Colen$^{*}$}%
\affiliation{James Franck Institute, University of Chicago, Chicago, IL 60637}
\affiliation{Department of Physics, University of Chicago, Chicago, IL 60637}
\affiliation{Joint Institute on Advanced Computing for Environmental Studies, Old Dominion University, Norfolk, Virginia 23539}
\author{Alexis Poncet$^{*}$}%
\affiliation{ Univ. Lyon, ENS de Lyon, Univ. Claude Bernard, CNRS, Laboratoire de Physique, F-69342, Lyon.}
\author{Denis Bartolo}%
\affiliation{ Univ. Lyon, ENS de Lyon, Univ. Claude Bernard, CNRS, Laboratoire de Physique, F-69342, Lyon.}
\author{Vincenzo Vitelli}
\affiliation{James Franck Institute, University of Chicago, Chicago, IL 60637}
\affiliation{Department of Physics, University of Chicago, Chicago, IL 60637}

\maketitle

\renewcommand{\theequation}{S\arabic{equation}}
\renewcommand{\thefigure}{S\arabic{figure}}
\renewcommand{\thetable}{S\arabic{table}}

\tableofcontents

\section{Neural network architecture and training}

The neural network used in this paper was implemented in Pytorch~\cite{paszke_pytorch_2019}. It consisted of a read-in network, a single-layer linear Fourier neural operator, and a read-out network.
The read-in and read-out networks were convolutional neural networks which learn nonlinear functions of their input fields. These used stacked ConvNext-style blocks which lifted the inputs to a high-dimensional feature space and learned the transformations in this representation. 
Each ConvNext block contained a grouped spatial convolution followed by two convolutions with kernel size 1 and an inverse bottleneck structure with expansion ratio 4. For example, a block with input and output channel size 16 had 3 convolutional layers: [ grouped $16\rightarrow16$, k1 $16\rightarrow64$, k1 $64\rightarrow16$]~\cite{liu_convnet_2022}. This structure enabled aggregation of short-range spatial information such as derivatives, and then subsequent local nonlinear transformations of those features. The read-in and read-out networks had identical structure, with two ConvNext blocks with channel sizes [1,64,64] and a final convolutional layer with channel size [64,1] and kernel size 1 to push the data back into the one-dimensional representation. These ConvNext blocks used 1D convolutions and acted only along the spatial dimension of the data. 

The linear neural operator can be interpreted as a Green's function $G(\mathbf{r})$ and can be represented numerically as a $d$-dimensional vector, where each entry sets the interaction strength at separation $\mathbf{r}$~\cite{gin_deepgreen_2021}. We choose an alternative representation as in~\cite{li_fourier_2021} and learn the Fourier-transformed Green's function $G(\mathbf{q})$. This reduces computational overhead for the convolution operation and enables measurement noise mitigation via a regularized loss function. 
\begin{align}
    \mathcal{L} = \underbrace{\sum_i \big( F_i - F^{NN}_i \big)^2 }_{\text{Reconstruction}} + \underbrace{\beta \sum_{\mathbf{q}} |G(\mathbf{q})|^2}_{\text{Regularization}}
    \label{eq:loss}
\end{align}
Such regularization improves interpretability by ensuring that the model learns a general and smooth solution to the PDE, rather than overfitting to noise appearing as high-wavenumber terms in $f(\mathbf{q})$. 

% 1. GFNN improves interpretability
In a standard deep convolutional neural network, such as a U-Net~\cite{ronneberger_u-net_2015}, each layer computes features from a local neighborhood of each pixel and the network propagates long-range information by stacking many convolutional layers.
The dual role of each layer in determining local and non-local features, as well as the number of layers needed to obtain a large receptive field, makes it difficult to interpret what deep networks have learned.

The presented architecture simplifies interpretation by separating the problem into three manageable pieces which can be analyzed separately.
The read-in and read-out modules compute local information at each point while the learned Green's function describes all of the long-range information propagation.
The network trains end-to-end, meaning a model that has learned to predict experimental data has also found a representation of the system that is governed by a PDE. Even if the inputs and outputs do not directly correspond to the fields in the continuum description, the read-in and read-out layers allow the network to learn a suitable transformation and a solution to the PDE governing those transformed variables. 
These features make this approach well-adapted for analyzing, testing, and identifying models of experiments with complex and noisy data. 

We used this neural network to analyze the dynamics of a one-dimensional microfluidic-particle stream which develops a shock in response to a jam induced at t = 0 and is predicted to be described by a Burger’s equation~\cite{champagne_stability_2011}. We coarse-grained the droplet positions by first performing background subtraction (with the background defined as the average image frame over the entire movie) and averaging in the transverse direction. We next convolved the movie with a Gaussian kernel with width $\sigma_t = 0.05$ s (3 frames sampled at 60 Hz) and $\sigma_x = 0.5$ mm $\approx 3$ particle diameters, and downsampled by a factor of 4 in space and time into a movie of size $7$ s $\times 18$ mm ($112\times200$ pixels). We segmented the resulting $\rho$ field into 6,200 overlapping $T'\times L' = 3$ s $\times 9$ mm ($50 \times 100$ pixel) segments to use as training examples. 
We trained the network using the Adam optimizer with learning rate $\lambda=10^{-4}$ on the convolutional layers and $\lambda=10^{-2}$ on the Green's function. We used a regularization parameter $\beta = 10^{-7}$ and trained for 1000 epochs using the Pytorch ReduceLROnPlateau scheduler (patience 10 epochs). 

\section{Trained network analysis}

We used the PySINDy sparse regression package~\cite{silva_pysindy_2020,kaptanoglu_pysindy_2022} to learn a dynamical rule for the machine-learned field $\phi(x, t)$ in Fig.~3. We used a library of linear spatial derivatives of the field $\phi$ up to fifth order and the sequentially-thresholded least-squares optimizer with $\tau=10^{-3}$ and $\alpha = 1$. The result was the equation (2) with the coefficients of $\partial_x^4,\,\partial_x^5$ thresholded to zero by the SINDy algorithm. 

We note that the coefficients are lower for the higher-order derivative terms -- this is a consequence of the derivatives $\partial_x^n \phi$ having larger magnitudes as $n$ increases. To quantify the relative importance of each term in (2), we multiply the absolute value of the learned coefficient by the L2 norm of the corresponding feature column in the SINDy library. This represents the typical weight that each term contributes to the dynamics $\partial_t \phi$. The results, reported in Table~\ref{tab:term_importance}, show the significance of the dispersive term $\partial_x^3$ as greater than that of the diffusive or damping terms.

\begin{table}[]
    \centering
    \begin{tabular}{c | c |c | c }
         \textbf{Term} &  \textbf{Coef.} & \textbf{L2 Norm} & \textbf{Importance} \\
         \hline 
         $\phi$ & -0.31 & $1.64 \times 10^2$ & 51.1 \\
         $\partial_x \phi$ & 1.86 & $1.56 \times 10^2$ & 290 \\
         $\partial_x^2 \phi$ & 0.003 & $9.97\times 10^2$ & 3.28 \\
         $\partial_x^3 \phi$ & -0.008 & $1.43\times 10^4$ & 113
    \end{tabular}
    \caption{Importance values for each term in the machine-learned equation (2). Importance is the product of the coefficient and the L2 Norm of the corresponding feature column in the SINDy feature library.
    }
    \label{tab:term_importance}
\end{table}

\begin{figure*}
    \centering
    \includegraphics{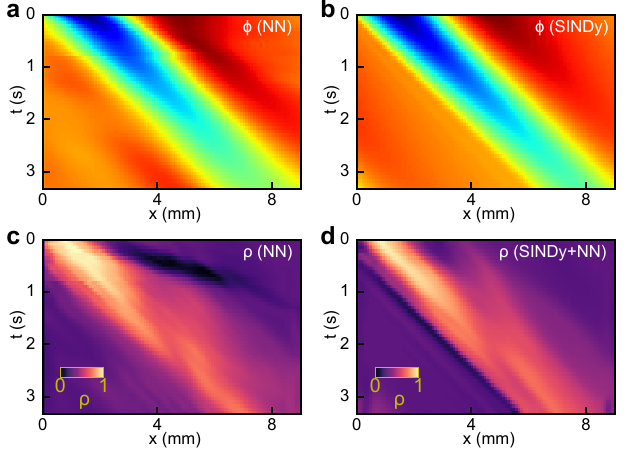}
    \caption{
        (\textbf{a}-\textbf{b}) Comparison of $\phi$ predicted by NN (\textbf{a}) and predicted by the learned equation (2) (\textbf{b}). The equation achieves an $R^2 = 0.96$.
        (\textbf{c}-\textbf{d}) Comparison of $\rho$ predicted by NN (\textbf{c}) and obtained by passing (\textbf{b}) through the NN read-out transformation (\textbf{d}). The mean absolute error between (\textbf{c}) and (\textbf{d}) is 0.063. The prediction in (\textbf{d}) achieves an MAE of 0.072 relative to experiments.
    }
    \label{fig:gfnn_internal}
\end{figure*}

To test the accuracy of (2), we used the read-in network to compute the initial condition $\phi(x, t=0)$ from the data. We simulated the above equation using a 4th order Runge-Kutta integration scheme over a $T = 3$ s window with $dt = 0.06$ s (both $T$ and $dt$ were the same as used by the neural network). This produced the $\phi_{\text{SINDy}}$ prediction shown in Fig.~\ref{fig:gfnn_internal}b. We then used the neural network read-out transformation to convert $\phi_{\text{SINDy}}$ to $\rho_{\text{SINDy}}$ shown in Fig.~\ref{fig:gfnn_internal}d. 

\begin{figure*}
    \centering
    \includegraphics{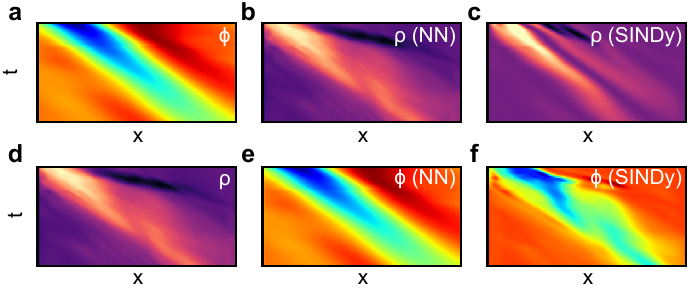}
    \caption{
        Learning the transformations $\rho \rightarrow \phi$ and $\phi \rightarrow \rho$ for microfluidics experiment. 
        (\textbf{a}) Plot of $\phi$ obtained via the read-in transformation and propagated using the neural operator layer.
        (\textbf{b}) Plot of $\rho$ predicted by by applying the read-out transformation to $\phi$.
        (\textbf{c}) $\rho_{\text{SINDy}}$ obtained by fitting the transformation $\phi \rightarrow \rho$ using the SINDy method. This method produces an equation with 14 terms and reaches an MAE of 0.085 relative to $\rho$ (NN) and an MAE of 0.100 relative to experiments.
        (\textbf{d}-\textbf{f}) Repeat of top row for learning the read-in transformation $\rho \rightarrow \phi$. Note that $\phi(x, t)$ is not predicted by integrating the PDE (2) as in Fig.~\ref{fig:gfnn_internal}, but is instead predicted by applying a learned equation to the experimental density field $\rho(x, t)$. The identified equation contains 23 non-zero coefficients and achieves an $R^2 = 0.70$ relative to $\phi$ (NN). 
    }
    \label{fig:burgerstransform}
\end{figure*}

We also attempted to learn rules for the read-in and read-out transformations in hopes that they might yield sparse representations similar to the neural operator. For each transformation $\phi \rightarrow \rho,\ \rho \rightarrow \phi$, we used a library of the input field, derivatives, nonlinear powers to order 3, $\sin$, and $\cos$. Our best learned models were not sparse. For the $\phi \rightarrow \rho$ transformation, we learned an equation which contained 14 terms and produced an MAE of 0.100 relative to experiments (MAE=0.085 relative to the $\rho$ field predicted by the neural network), see Fig.~\ref{fig:burgerstransform}a-c). For the $\rho \rightarrow \phi$, we learned an equation which contained 23 terms and achieved an $R^2 = 0.70$ relative to the $\phi$ field predicted by the NN, see SI Fig.~\ref{fig:burgerstransform}d-f).
\iffalse
\begin{widetext}
    \begin{multline}
        \rho = 
        \left( -0.114 \, \phi_x + 0.027 \, \phi \, \phi_x^2 + 1.828 \, \phi_x^3 \right) \sin \phi + 
        \left( -0.001 \, \phi_x^2 - 0.465\, \phi_x^3 \right) \cos \phi + \\
        \left( -2390 - 0.700 \, \phi\, \phi_x + 805\, \phi_x^2 - 0.346\, \phi \, \phi_x^2 - 37.4\, \phi_x^3 \right) \sin \phi_x + 
        \left( 0.339 + 2390\, \phi_x - 0.052\, \phi \, \phi_x + 6.01\,  \phi_x^2 \right) \cos \phi_x
        \label{eq:phi2rho}
    \end{multline}
    \begin{multline}
    \phi = 
    \left( -2110 - 18400\, \rho_x - 209\, \rho^2 - 30.1\, \rho^3 \right) \cos \rho + \\
    \left( 68100\, \rho^2 + 1430\, \rho^3 \right) \sin \rho_x + 
    \left( 2110 + 18500\, \rho_x - 848\, \rho^2 - 78200 \, \rho^2\, \rho_x \right) \cos \rho_x
    \label{eq:rho2phi}
    \end{multline}
\end{widetext}
\fi

\section{Linearizing Burgers' equation}
\label{sec:colehopf}

Burgers' equation is a paradigmatic example describing shock generation in fluids. 
Consider an initial value problem for the viscous Burger's equation.
\begin{align}
    \partial_t u + u \partial_x u &= \nu \partial_x^2 u \label{eq:burgers} \\
    u(x, t=0) &= f(x)
\end{align}
On the surface, this is a nonlinear PDE and thus is not tractable using Green's function methods. However, it can be mapped onto a linear problem via a Cole-Hopf transformation.
\begin{align}
    u = -2 \nu \frac{1}{\phi} \partial_x \phi
    \label{eq:colehopf}
\end{align}
This change of variables turns (\ref{eq:burgers}) into a linear diffusion equation
\begin{align}
    \partial_t \phi - \nu \partial_x^2 \phi = 0  \label{eq:diffusion}
\end{align}
\begin{align}
    \phi(x, t=0) = \exp \bigg( -\frac{1}{2\nu} \int f(y)\,dy \bigg)
\end{align}
This analogous initial value problem is equivalent to the response to an instantaneous impulse $f(x)$ at $t = 0$, and thus it can be solved by integration with the Green's function for the one-dimensional diffusion equation
\begin{align}
    G(x, t) = \frac{1}{\sqrt{4\pi\nu t}} \exp \bigg( -\frac{x^2}{4 \nu t} \bigg) \Theta(t)
    \label{eq:heatkernel}
\end{align}
The Cole-Hopf result is a standard example of how nonlinear systems can become tractable to Green's function methods via a suitable transformation. 

\iffalse
\section{KdV-Burgers' equation}

To further analyze what was learned by the neural network trained on microfluidic-particle experiments, we turned to numerical investigations of a general Korteweg-de Vries-Burgers' equation
\begin{equation}
    \partial_t u + a\, u \partial_x u - D \partial_x^2 u + b\, \partial_x^3 u = 0
    \label{eq:kdvburgers}
\end{equation}
The $b = 0$ case is Burgers' equation (\ref{eq:burgers}). 
The $D = 0$ case is the Korteweg-de Vries (KdV) equation for shallow water waves. This equation admits exact soliton solutions of the form~\cite{wang_exact_1996}
\begin{equation}
    u(x, t) = \frac{c}{2} \sech^2 \left[ \frac{\sqrt{c}}{2} (x - c\, t - x_0) \right]
    \label{eq:soliton}
\end{equation}
The system's response to an initial perturbation is to split into a superposition of solitons moving at different speeds~\cite{zabusky_interaction_1965}. 
The full KdV-Burgers equation has a traveling wave solution which is a superposition of a symmetric wave similar to (\ref{eq:soliton}) and an asymmetric kink~\cite{wang_exact_1996}
\begin{multline}
    u(x, t) = \frac{3 D^2}{25 a b} \bigg[ \sech^2 \frac{D}{10 b} \left(x - \frac{6 D^2}{25}\, t \right) - \\
    2 \tanh \frac{D}{10 b} \left(x - \frac{6 D^2}{25}\, t \right) - 2 \bigg] 
    \label{eq:solitonKdVB}
\end{multline}
\fi

\section{Applying KdV-Burgers' to experiments}

\begin{figure}
    \centering
    \includegraphics[width=0.5\textwidth]{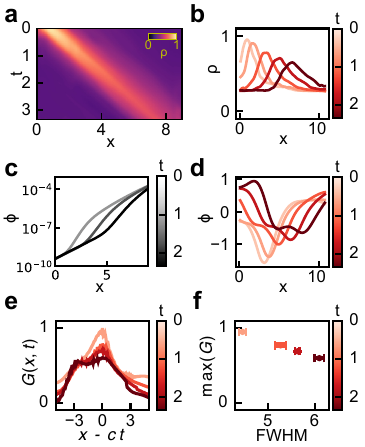}
    \caption{Neural operator learns to predict density field that obeys Burgers' equation.
        (\textbf{a}) Neural network trained on Burgers' data in Fig.~3b predicts diffusive spreading of initial jam.
        (\textbf{b}) Neural network predictions of the density profile $\rho(x, t)$ at different time points.
        (\textbf{c}) Cole-Hopf transformation of the density field into an intermediate variable.
        (\textbf{d}) The intermediate variable $\phi(x, t)$ predicted by the neural operator is distinct from the Cole-Hopf variable in (\textbf{c}).
        (\textbf{e}) Neural network Green's function in the comoving frame at different time points.
        (\textbf{f}) The neural network's learned Green's function decreases in magnitude and increases in width over time.
    }
    \label{fig:dedalusburgers}
\end{figure}

In Figure~\ref{fig:dedalusburgers} we compared simulations of a Burgers' equation, a KdV equation, and a KdV-Burgers' equation from experimental initial conditions. 
We used a physics-informed neural network (PINN)~\cite{raissi_physics-informed_2019} to fit the coefficients of each equation, as we found other approaches had trouble fitting the $\partial_x^3 \rho$ term, likely due to its amplification of experimental noise.  Each PINN was each fully-connected neural network with 5 hidden layers of 50 neurons with sine activations and trained on 10000 randomly selected points from the dataset. They minimized the joint loss function
\begin{align}
    \mathcal{L} = \frac{1}{N}\sum_{i=1}^N \big[\rho_i - \hat{\rho}_i\big]^2 + 
    \frac{\beta}{N} \sum_{i=1}^N \big[ \partial_t \rho_i +
     (c - \alpha)\rho_i \partial_x \rho_i - D \partial_x^2 \rho_i - \beta \partial_x^3 \rho_i \big]^2
\end{align}
The first term is a reconstruction loss using the mean-squared deviation between prediction and experiments. The second term is a physics-informed loss which penalizes the deviation from the prescribed equation (3). 
We used $\beta = 10$ to weigh the physics-informed loss more heavily than the reconstruction loss. 
Following~\cite{raissi_physics-informed_2019}, we minimized the joint objective function using the L-BFGS optimizer. The final components of the loss function for each equation are summarized in Table~\ref{tab:pinn_results}.

\begin{table}[ht]
    \centering
    \begin{tabular}{c | c |c}
         \textbf{Equation} &  \textbf{MSE Loss} & \textbf{Phys. Loss} \\
         \hline
         Burgers' & $3.03 \times 10^{-3}$ & $1.66 \times 10^{-5}$ \\
         KdV & $3.48 \times 10^{-3}$ & $2.91 \times 10^{-5}$ \\
         KdV-Burgers\' & $\mathbf{ 2.98 \times 10^{-3}}$ & $\mathbf{1.37 \times 10^{-5}}$
    \end{tabular}
    \caption{PINN loss terms for variants of the general KdV-Burgers' equation. The full equation (3) achieves the lowest reconstruction error and the lowest physics error.}
    \label{tab:pinn_results}
\end{table}

The Burgers' and KdV-Burgers' PINNs achieved similar reconstruction accuracy, while the KdV PINN performed worse. By construction, PINNs allow a tradeoff between faithfulness to the data and adherence to the prescribed physics. We observe that the KdV-Burgers' PINN achieves a lower physics-informed loss than the other two models. The combination of low reconstruction error while also closely satisfying the physical constraint suggests that the KdV-Burgers' equation may be more suitable to the experimental data than either the Burgers' or KdV limits.

To further test the accuracy of the PINN-identified coefficients, we integrated each equation from experimental initial conditions (Fig.~3b). These results agreed with the PINNs, with Burgers' and KdV-Burgers' performing well and KdV achieving much lower accuracy. The KdV-Burgers' equation achieved higher accuracy near the shock front (Fig.~3c, Table~\ref{tab:error_rates}). Together, the PINN results and Fig.~3 corroborate the conclusion that the neural operator learned KdV-like physics in order to preserve the shock sharpness over time.

\begin{table}[]
    \centering
    \begin{tabular}{c | c |c}
         \textbf{Solution} &  \textbf{MAE} & \textbf{Shock-MAE} \\
         \hline \hline
         Neural network &  0.049 & 0.051 \\
         \hline
         Burgers' & 0.067 & 0.074 \\
         KdV & 0.11 & 0.14 \\
         KdV-Burgers\' & 0.068 & 0.067 \\
         \hline 
         Baseline & 0.17 & 0.34
    \end{tabular}
    %No acronym
    \caption{ Error rates for neural network (Fig.~2) and analytic solutions (Fig.~3). We compute the mean absolute error (MAE) $\langle |\rho - \rho_0| \rangle$ for the whole field and within a 1-mm region centered at the shock (Shock-MAE). Baseline reports the error rates of predicting a uniform density field given by the mean of the movie.
    }
    \label{tab:error_rates}
\end{table}

\section{Mapping the microscopic model onto the KdV-Burgers equation}
In this section, we derive a PDE that  describes the long wavelength dynamics of our microscopic model. 
The method follows the derivation of the KdV equation from the standard microscopic model offered by the Fermi-Pasta-Ulam-Tsingou chain, see \cite{Dauxois2006} for a thorough presentation.

\subsection{Derivation of the KdV-Burgers equation} \label{ss:derivKdV}
We start from the microscopic equations of motion (Eq.~(4) of the main text),
\begin{equation}\label{eq:micro_eq0}
    \zeta(\partial_t R_n-v_0) = \sum_{m\neq n}\left( f_{m\to n}^H + f_{m\to n}^E \right).
\end{equation}
In the following we set $\zeta = 1$ without loss of generality. We define the local density $\rho(x)$ as
\begin{equation} \label{eq:def_rho}
    \rho(x=na) = \frac{1}{R_{n+1}-R_n-a}
\end{equation}
where $a$ is the lattice spacing, and $x$ is the continuous analog of $n$. Our goal is to derive the dynamical equation obeyed by $\rho(x,t)$ in the limit of small density fluctuations and long wavelengths.

We start from a reference configuration on the infinite line with uniform lattice spacing $a$, so that the reference positions of the particles are $R_n^0 = na$. 
In term of the displacements $u_n(t) = R_n(t) - R_n^0$, Eq.~\eqref{eq:micro_eq0} becomes
\begin{equation} \label{eq:micro_eq1}
 \partial_t u_n = \sum_{j\geq 1} \left[ f_H(ja + u_{n+j} - u_n) + f_H(j a + u_n - u_{n-j})\right] + \sum_{j\geq 1} \left[f_E(ja + u_{n+j} - u_n) - f_E(j a + u_n - u_{n-j})\right],
\end{equation}
where $f_H(r)$ is the distance-dependent hydrodynamic force . In the narrow channels of our experiments, the hydrodynamic interactions are exponentially screened. 
$f_E(r)$ is the elastic force.
Within a  linear spring approximation, we have $f_E(a+\Delta r) = k\Delta r$, and the sum on $j\geq 1$ reduces to the $j=1$ term only. We stress that the elastic interactions are reciprocal, in the sense that they obey Newton's third law of motion (action-reaction principle).
Reciprocity manifests itself by the minus sign difference between the two 
forces in the second sum of Eq.~\eqref{eq:micro_eq1}
By contrast, the hydrodynamic interactions are fully non-reciprocal, they do not obey the action-reaction principle, which results in a $+$ sign in the first sum of Eq.~\eqref{eq:micro_eq1}.
%In the following, we set the friction coefficient $\zeta=1$.
%$\zeta = 1$, only term $k$ if nearest neighbors only. Fast decay.

When the deformations ($u_{n\pm j} - u_n$) are small, we can expand $f_H$ and $f_E$ up to second order,
\begin{multline}\label{eq:micro_eq2}
    \partial_t u_n = \sum_{j\geq 1} \left[ 2f_H(ja) + f_H'(ja) (u_{n+j} - u_{n-j}) + \frac{f_H''(ja)}{2}\left((u_{n+j}- u_n)^2 + (u_n - u_{n-j})^2 \right) + \dots  \right] \\
    + \sum_{j\geq 1} \left[ f_E'(ja) (u_{n+j} + u_{n-j} - 2u_n) + \frac{f_E''(ja)}{2}\left((u_{n+j}- u_n)^2 - (u_n - u_{n-j})^2 \right) + \dots \right].
\end{multline}
To build a continuous limit, we consider deformations of long wavelength, i.e of order $a/\epsilon$ with $\epsilon\to 0$. 
We can then define a (dimensionless) continuous spatial variable $\tilde x = \epsilon\frac{x}{a} = \epsilon n$. We then define the displacement field as $u_n = \nu(\epsilon) a \tilde u(\tilde x)$. 
$\nu(\epsilon)$ is a scale factor that defines the magnitude of the displacements described by our continuum theory.  
The magnitude of the  linear  deformations are then of order  $\nu \epsilon$. 
More precisely, we can expand  the local deformations in powers of $\epsilon$ as 
\begin{equation}
    u_{n\pm j} - u_n = \nu a \left[\tilde u(\tilde x\pm j\epsilon) - \tilde u(\tilde x)\right] = \nu a \left[\pm (j\epsilon) \partial_{\tilde x} \tilde u + \frac{(j\epsilon)^2}{2} \partial_{\tilde x\tilde x}\tilde u\pm \frac{(j\epsilon)^3}{6} \partial_{\tilde x\tilde x\tilde x}\tilde u + O(\epsilon^4)\right].
\end{equation}
Injecting this expansion in Eq.~\eqref{eq:micro_eq2}, we find that $\tilde u$ obeys
\begin{equation} \label{eq:macro_eq1}
    \partial_t \tilde u(\tilde x) = \left[V - \epsilon c\partial_{\tilde x} \tilde u - \epsilon^2 \nu \frac{\alpha}{2} (\partial_{\tilde x} \tilde u)^2 + \epsilon^3 \beta \partial_{\tilde x\tilde x\tilde x} \tilde u \right] + \left[\epsilon^2 K_1 \partial_{\tilde x\tilde x}\tilde  u + \epsilon^3 \nu K_2 (\partial_{\tilde x} \tilde u) (\partial_{\tilde x\tilde x} \tilde u)\right] + O(\epsilon^4),
\end{equation}
where the parameters $V, c, \alpha, \beta, K_1, K_2$ are defined as
\begin{align}
    V &= 2\sum_{j\geq 1} f_H(ja), & c &= -2\sum_{j\geq 1} j f_H'(ja), & \alpha &= -2a\sum_{j\geq 1} j^2 f_H'(ja), \\
    \beta &= \frac{1}{3}\sum_{j\geq 1} j^3 f_H'''(j\epsilon), & K_1 &= \sum_{j\geq 1} j^2 f_E'(ja), & K_2 &= \frac{a}{2}\sum_{j\geq 1}j^3 f_E''(ja).
\end{align}
We note that in the case of nearest neighbor interactions, only the term $j=1$ remains. In particular, for linear springs, we have $K_1=k$ and $K_2=0$.

We can now write an equation of motion for the local deformations
, and disregard the advection term ($\partial_{\tilde x} \tilde u$), as it  can be eliminated by a change of reference frame.
In all that follows, we use the new coordinate $\tilde y=\tilde x-\epsilon c t= \epsilon(n-ct)$.
The equation for the deformation field $\tilde w(\tilde y) = \partial_{\tilde y} \tilde u$ then reduces to
\begin{equation} \label{eq:macro_eq2}
    \partial_t \tilde w(\tilde y) = \epsilon^2 K_1\partial_{\tilde y\tilde y} \tilde w - (\epsilon^2 \nu) \alpha \tilde w\partial_{\tilde y} \tilde w + \epsilon^3 \beta \partial_{\tilde y\tilde y\tilde y} \tilde w + (\epsilon^3 \nu) K_2 \partial_{\tilde y}(\tilde w\partial_{\tilde y} \tilde w)  + O(\epsilon^4).
\end{equation}
We note that the $K_2$ term that originates from the elastic interactions is of order $\epsilon^3\nu$.
It is therefore negligible comparated to   the non-linear correction that originates  from the hydrodynamic interactions (of order $\epsilon^2\nu$).
In other words, without loss of generality, we the linear-spring approximation is sufficient to explain the large-scale physics of our system.

The non-reciprocal non-linear term ($\tilde w\partial_{\tilde y}\tilde w$) is known to yield shocks. The question is how are these shocks regularized. 
When the elastic interactions are strong ($K_1\sim 1$), the elasticity regularizes perturbations of order $\nu\sim 1$, leading to the Burgers equation (at order $\epsilon^2$)
\begin{equation}
    \partial_{\tilde t'} \tilde w(\tilde y) = K_1\partial_{\tilde y\tilde y} \tilde w -\alpha \tilde w\partial_{\tilde y} \tilde w,
\end{equation}
with the rescaled time $\tilde t'=\epsilon^2 \tau$.

When elastic interactions are vanishingly small ($K_1=K_2=0$), the shocks are regularized by the dispersive term induced by the non-reciprocal interactions. 
For perturbations of order $\nu\sim \epsilon$, we find the well-known KdV equation (at order $\epsilon^3$),
\begin{equation}
    \partial_{\tilde t} \tilde w = -\alpha \tilde w\partial_{\tilde x} \tilde \tilde w + \beta \partial_{\tilde x\tilde x\tilde x} \tilde w,
\end{equation}
with the rescaled time $\tilde t = \epsilon^3 t$.

Finally, when the elastic interactions are weak, $K_1=\epsilon \tilde K_1$ with $\tilde K_1$ of order $1$, the dispersion and the damping both contribute to regularize shocks. 
Eq. \eqref{eq:macro_eq2} then takes the generic form of a KdV-Burgers equation (at order $\epsilon^3$),
\begin{equation}
    \partial_{\tilde t} \tilde w = \tilde K_1 \partial_{\tilde y\tilde y} \tilde w -\alpha \tilde w \partial_{\tilde y} \tilde w + \beta \partial_{\tilde y\tilde y\tilde y} \tilde w.
\end{equation}
To  compare the prediction of our continuum theory with microscopic simulations and experiments,  we  recast this dynamics into an equation of motion for the density field. 
For small density fluctuations $\rho - a^{-1} = \epsilon^2\tilde \rho$ (see Eq. \eqref{eq:def_rho}), and
noticing that $\tilde\rho = -\tilde w$, we find
\begin{equation} \label{eq:kdvb1}
    \partial_{\tilde t} \tilde\rho = \tilde K_1 \partial_{\tilde y\tilde y} \tilde\rho + \alpha \tilde\rho \partial_{\tilde y} \tilde\rho + \beta \partial_{\tilde y\tilde y\tilde y}.\tilde\rho,
\end{equation}
We stress that the scaling of all physical quantities with $\epsilon$ was essential to control our perturbative analysis. However, none of our prediction explicitly depend on $\epsilon$. We can come back to the original variables, $t=\epsilon^{-3}\tilde t$, $x = \epsilon^{-1}\tilde y + ct$, $K_1 = \epsilon^{-1}\tilde K_1$, $\rho = a^{-1} + \epsilon^2\tilde \rho$ and obtain
\begin{equation} \label{eq:kdvb2}
    \partial_t\rho = -c\partial_x \rho + K_1 \partial_{xx} \rho + \alpha \rho \partial_x\rho + \beta \partial_{xxx} \rho,
\end{equation}
which corresponds to Eq.~(3) of the main text with $K_1 = D$. This equation is a priori valid for small pertubations, but we show in Fig. \ref{fig:simuOthers} that it gives reasonable predictions even for large perturbations.
%with $\alpha = -2A$ and $\beta = B$.

In practice, in our numerical simulations, we choose $f_H(r) = -\left(\frac{\pi}{W}\right)^2 \mathrm{csch}^2\left(\frac{\pi r}{W}\right)$ with $W=2$ (and $a=1$). This yields the following values for the parameters: $c=-3.44, \alpha = -13.4, \beta=0.713$.

\subsection{Some properties of the KdV and KdV-Burgers equation}
In this section we recall some know properties of the KdV and KdV-Burgers equations~\cite{Dauxois2006,Polyanin2003}.

Let us first consider the KdV equation,
\begin{equation}
    \partial_t \rho = \alpha \rho \partial_x\rho + \beta \partial_{xxx} \rho.
\end{equation}
It is an integrable system, and its solution (given an initial condition) can be derived analytically by the inverse scattering method. In general, a solution is the sum of solitons propagating in one direction (to the left in Fig. \ref{fig:simuOthers}), and dispersive waves that move in the opposite direction (to the right in Fig. \ref{fig:simuOthers}).
%Let us remind the reader of the one-soliton and two-soliton solutions of the KdV equation.
For some specific initial conditions, the solutions reduce to a superposition of solitons only.
In particular, the one-soliton solution of the KdV equation is parametrized by a single parameter $\kappa >0$ that sets the width, the amplitude and the speed of the soliton:
\begin{equation}
    \rho(x, t) = \frac{12\beta\kappa^2}{\alpha}\mathrm{sech}^2\left[\kappa \left(x+4 \beta \kappa^2 t \right)\right].
\end{equation}
The soliton propagates at velocity $-4\beta\kappa^2$. 
In our case $\beta >0$, solitons propagate to the left.

The exact two-soliton solution of the KdV equation is governed by two parameters $\kappa_1, \kappa_2 >0$ setting the speeds of the solitons (when they are well separated):
\begin{gather} \label{eq:kdv_twosolitons}
    \rho(x, t) = \frac{12\beta}{\alpha} \frac{\kappa_1^2 - \kappa_2^2}{\left(\kappa_1 \coth X_1 - \kappa_2\tanh X_2\right)^2} \left(\frac{\kappa_1^2}{\sinh^2 X_1} + \frac{\kappa_2^2}{\sinh^2 X_2}\right) \\
    X_n = \kappa_n \left(x+4\beta\kappa_n^2 t\right),
\end{gather}
where $t=0$ is the time at which the two solitons cross one another. 
We use this solution to initialize our numerical simulations. By noting that $\rho - \rho_0 \sim -\partial_x u$, we find
\begin{equation} \label{eq:displ2solitons}
    u(x, t) = -\frac{12\beta}{\alpha} \frac{\kappa_1^2 - \kappa_2^2}{\kappa_1 \coth X_1 - \kappa_2 \tanh X_2}.
\end{equation}
We use this result to initialize our discrete chain in the simulations presented in Fig.~4 of the main text.
In passing, we note that one can build an exact $N$-soliton solution of the KdV equation~\cite{Polyanin2003}.

The KdV-Burgers equation, Eq. \eqref{eq:kdvb1}, is no longer an integrable system. 
We choose to solve it by numerical integration (using a spectral solver described below). 
%We nevertheless note that some exact solutions exist~\cite{Wang1996}, and that the behavior of damped solitons can be investigated perturbatively~\cite{Kivshar1989}.

\subsection{Cubic non-linearities and modified KdV-Burgers equation}
We make a technical remark that allows us to put our results in a broader context.
In principle the expansion of the non-reciprocal interactions could solely include odd order terms:
\begin{equation}
    f_H(ja + u_{n+j}-u_n) = f'_H(ja) (u_{n+j} - u_n) + \frac{f_H'''(ja)}{6}  (u_{n+j} - u_n)^3 + \dots
\end{equation}
In this case, Eqs.~\eqref{eq:macro_eq1} and \eqref{eq:macro_eq2} would become
\begin{align}
    \partial_t \tilde u(\tilde x) &= \left[- \epsilon c\partial_{\tilde x} \tilde u + \epsilon^3 \nu^2 \frac{\gamma}{3} (\partial_{\tilde x} \tilde u)^3 + \epsilon^3 \beta \partial_{\tilde x\tilde x\tilde x} \tilde u \right] + \left[\epsilon^2 K_1 \partial_{\tilde x\tilde x}\tilde  u + \epsilon^3 \nu K_2 (\partial_{\tilde x} \tilde u) (\partial_{\tilde x\tilde x} \tilde u)\right] + O(\epsilon^4), \\
    \partial_t \tilde w(\tilde y) &= \epsilon^2 K_1\partial_{\tilde y\tilde y} \tilde w + (\epsilon^3 \nu^2) \gamma \tilde w^2\partial_{\tilde y} \tilde w + \epsilon^3 \beta \partial_{\tilde y\tilde y\tilde y} \tilde w + (\epsilon^3 \nu) K_2 \partial_{\tilde y}(\tilde w\partial_{\tilde y} \tilde w)  + O(\epsilon^4),
\end{align}
with $\gamma=3a^2\sum_{j\geq 1} j^3 f_H'(ja)$. When the elasticity is small $K_1 = \epsilon\tilde K_1, K_2 = \epsilon \tilde K_2$, we may take $\nu\sim 1$ and obtain the following equation at order $\epsilon^3$,
\begin{equation}
    \partial_{\tilde t} \tilde w(\tilde y) =  K_1\partial_{\tilde y\tilde y} \tilde w + \gamma \tilde w^2\partial_{\tilde y} \tilde w + \beta \partial_{\tilde y\tilde y\tilde y} \tilde w,
\end{equation}
with $\tilde t = \epsilon^3 t$. The same equation holds for the scaled density fluctuations $\tilde \rho$. This PDE is known as the modified KdV-Burgers equation, which reduces to the mKdV (modified Korteweg-de Vries) equation when there is no elasticity~\cite{Dauxois2006,Polyanin2003}. Like the KdV equation, the mKdV equation is integrable and enjoys exact soliton solutions.

\subsection{Discrete chain and Volterra lattice}
Deriving continuous descriptions as we did in the previous sections is appealing, but we note for completeness that some results can already be obtained at the discrete level for a specific choice of forces. Let us consider exponentially decaying non-reciprocal forces
\begin{equation}
    f_H(r) = A e^{-|r|/b},
\end{equation}
with no potential forces $f_E = 0$. In this case, Eq.~\eqref{eq:micro_eq0} (with $v_0=0$ for simplicity) reads
\begin{equation}
    \partial_t R_n = \tilde A \left[e^{-\Delta_n/b} + e^{-\Delta_{n-1}/b} \right],
\end{equation}
where $\tilde A = A e^{-a}/\zeta$ and the local deformation is $\Delta_n = R_{n+1}-R_n-a$, with $R_{n+1}-R_n>0$ at all times.
This equation takes a simple form when written in terms of the quantity $v_n = -({b}/{\tilde A}) e^{-\Delta_n/b}$,
\begin{equation}
    \partial_t v_n = v_n(v_{n+1} - v_{n-1}).
\end{equation}
This set of equations is known as the Volterra lattice, as it describes a minimal predators and preys model involving an infinite number of species.

The Volterra lattice is an integrable system. We can define a Lax pair $(M, L)$ satisfying $\partial_t L = [M, L]$, where $L$ and $M$ are operators acting on sequences $f_n$ as $(Lf)_n = v_n f_{n-1} + f_{n+1}$ and $(Mf)_n = (v_n + v_{n+1})f_n + f_{n+2}$.
Importantly, the Volterra lattice enjoys solitonic solutions~\cite{Adler2019}.

We note here the strong similarity between the Volterra lattice and the Toda chain. 
The Toda chain is an integrable discrete model akin to the Fermi-Pasta-Ulam-Tsingou (FPUT) chain~\cite{Toda1967}.
%It has been argued~\cite{???} that the Toda chain is an even better model of the FPUT chain than the KdV equation.
From this perspective, instead of thinking of the non-linear waves seen in our simulations as KdV solitons, we may alternatively see them as (approximate) excitations of a Volterra lattice for suitably chosen parameters.

%%%%%
\section{Numerical simulations}

\subsection{Numerical solution of the microscopic equations}
We discuss the details of the numerical simulations reported in Fig.~4 (Main text) and Fig.~\ref{fig:simuOthers} (below).
We initially set $N=500$ particles on a line at positions $R_n(t=0) = i + u_n^0$ for $i=n,\dots, N$ where we used a lattice spacing $a=1$ and $u_n^0$ is the initial perturbation. Periodic boundary conditions are enforced, in a box length $L = N + u_N - u_1$, so that there is no deformation at the edge of the box.
In the main text (Fig.~4), we input an exact two-solition solution of the KdV equation: $u_n^0 = u(n a, t_0)$ using Eq.~\eqref{eq:displ2solitons} with $t_0 = -10^3$ and the soliton speeds $V_1 = 4\beta\kappa_1^2 = 0.05$ and $V_2 = 4\beta\kappa_2^2 = 0.01$. 
The parameters $\alpha$ and $\beta$ are chosen according to the derivation of subsection \ref{ss:derivKdV}.

In Fig.~\ref{fig:simuOthers}, we chose an ``arbitrary'' intial condition as the sum of two gaussians:
\begin{equation}
    \rho^0(x) - a^{-1} \approx h_1 e^{-\frac{(x-x_1)^2}{2w_1^2}} + h_2 e^{-\frac{(x-x_1)^2}{2w_1^2}}.
\end{equation}
More precisely, we used
\begin{equation}
    u_n^0 = -\sqrt\frac{\pi}{2}\left[w_1h_1\erf\left(\frac{n-x_1}{\sqrt{2}w_1}\right) + w_2h_2\erf\left(\frac{n-x_2}{\sqrt{2}w_2}\right) \right],
\end{equation}
with $x_1-x_2 = 30$, $w_1 = w_2 = 10$. In Fig.~\ref{fig:simuOthers}a, $(h_1, h_2)=(0.015, -0.005)$ while in Fig.~\ref{fig:simuOthers}b, $(h_1, h_2)=(0.15, -0.02)$. 
The $x$ axes of Fig.~4 and Fig.~\ref{fig:simuOthers} are shifted so that the perturbation is initially around the origin.

The equations of motion are
\begin{equation} \label{eq:sim}
    \partial_t u_n = \sum_{n\neq m} f_H(d_{n, m}) + k(u_{n+1} + u_{n-1} - 2 u_n)
\end{equation}
where $d_{n,m}$ is the distance between particles $n$ and $m$ (for $m>n$, $d_{n,m} = \mathrm{per}(m-n+u_m-u_n)$ where $\mathrm{per}$ periodizes the distance in the interval $(-L/2, L/2]$). The hydrodynamic force is
$f_H(r) = -\left(\frac{\pi}{W}\right)^2 \mathrm{csch}^2\left(\frac{\pi r}{W}\right)$ with $W=2$ (see subsection \ref{ss:hydro}). The elastic constant is $k=0, 0.02$ or $0.2$ depending on the panels of the figures.

The $N$ coupled equations of motion \eqref{eq:sim} are integrated using the Julia package \texttt{DifferentialEquations.jl}~\cite{Rackauckas2017} with algorithm \texttt{AutoVern7(Tsit5())} and maximal timestep $0.1$.

\subsection{Additional numerical simulations}
In Fig. \ref{fig:simuOthers}, we present numerical simulations with two other initial conditions in addition to the ones shown in the main text. 
Fig. \ref{fig:simuOthers}a, shows an generic small perturbation that decays into two solitons traveling to the left, and dispersive waves travelling to the right. The agreement with the KdV-Burgers equation is quantitative and the strong elasticity limit is described by Burgers' equation.

Fig. \ref{fig:simuOthers}b shows a perturbation of larger amplitude. In the absence of elasticity, a large number of solitons is generated in agreement with the KdV prediction. While the perturbation is rather large, it is still well described the KdV-Burgers equation (and the Burgers equation at large damping).

\subsection{Integration of KdV, KdV-Burgers and Burgers equations}
We now explain how we solve the nonlinear PDEs (KdV, Burgers and KdV-Burgers) to compare our continuum models to the discrete numerical simulations.

\textbf{KdV equation.} The dashed gray curves in Fig.~4e correspond to the analytical 2-soliton solution of the KdV equation, Eq. \eqref{eq:kdv_twosolitons}, at times $t_0 + 2\cdot 10^3$ and $t_0 + 4\cdot 10^3$.

\begin{figure*}
    \centering\includegraphics{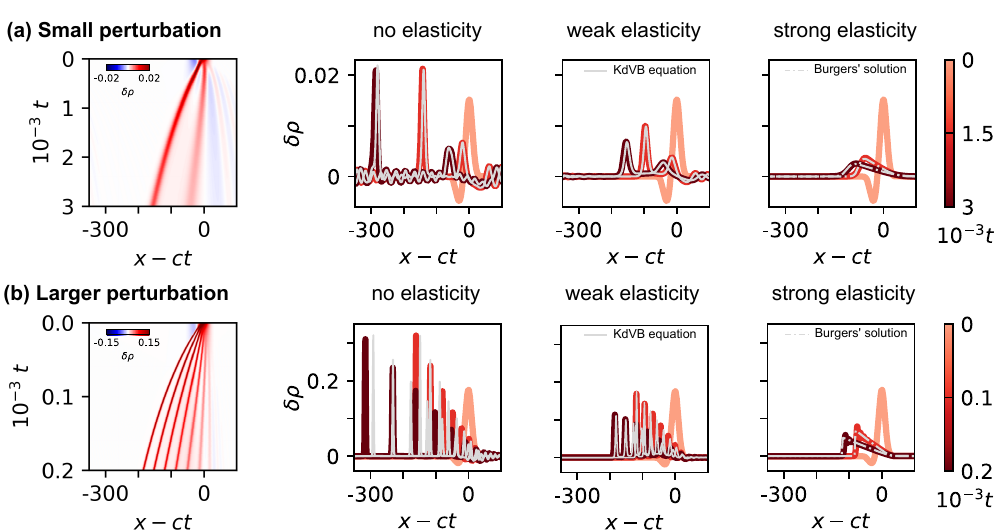}
    \caption{Numerical simulations with different intial conditions. (a) Small perturbation ($\delta\rho\sim 0.2$). The left panel is the spatiotemporal plot corresponding to a weak elasticity ($k=0.02$). The initial condition decomposes into two solitons and dispersive waves as expected from the KdV phenomenology. The three other panels correspond to no elasticity ($k=0$), weak elasticity ($k=0.02$) and strong elasticity ($k=0.2$). In the three cases the evolution is well described by the KdV-Burgers equation derived from the microscopic parameters (thin gray line), see text (without elasticity, this reduces to the KdV equation). The large elasticity case is well described by the Burgers equation (dashed lines). (b) Larger perturbation ($\delta\rho\sim 0.15$). The four panels correspond to the same parameters as (a). In this case, the initial condition decays in a large number of solitons, well predicted by the KdV-Burgers equation. 
    The strong elasticity limit is still predicted by the Burgers equation.}
    \label{fig:simuOthers}
\end{figure*}

\textbf{KdV-B equation.} The continuous gray curves in Fig.~4d and Fig.~\ref{fig:simuOthers} correspond to a numerical integration of the KdV-Burgers equation~\eqref{eq:kdvb2} with the parameters derived in subsection~\ref{ss:derivKdV}. 
We use the Dedalus package (spectral solver)~\cite{Burns2020} with standard parameters.

\textbf{Burgers equation.} The dashdotted gray line in Fig.~4e and Fig. \ref{fig:simuOthers} is the solution of the Burgers equation, with the parameters of subsection \ref{ss:derivKdV}. We use the results of section \ref{sec:colehopf} to compute the time evolution of the initial condition: Cole-Hopf transformation, and (numerical) convolution with the Green's function of the diffusion equation.

\subsection{Hydrodynamic interactions} \label{ss:hydro}
In this last section, we provide more details on the specifics of the hydrodynamic interactions relevant to model our experiments.
 More details can be found in Ref. \cite{Beatus2007}. 
 When squeezed by the confining walls, the droplets do not act as passive tracers.
 The viscous friction with the walls reduce their advection speed. We can therefore think of the droplets as  intruders driven through a confined viscous fluid. 
 Their motion perturb the mean flow and induce a velocity field $\vv(x, y)$ that contribute to the advection of  the other droplets.

We consider the 2D Stokes flow $\vv_1(x, y)$ around a droplet in infinite space. It is incompressible and satisfies $\nabla\cdot\vv_1(\rr) = 2\pi \sigma \hat\ee_x \cdot \nabla \delta(\rr)$ (source dipole). The solution can be expressed either in term of a potential $\phi_1(x, y)$ or a streamfunction $\psi_1(x, y)$. The complex potential defined as $w_1(z)=\phi_1(z) + i\psi_1(z)$, and the complex velocity $v_1(z) = v_{1,x}(z) - i v_{1,y}(z)$ are related by the formula $v(z) = \frac{\partial w}{\partial z}$. Their expressions are
\begin{align}
    w_1(z) &= \frac{\sigma}{z}, & 
    v_1(z) &= -\frac{\sigma}{z^2}.
\end{align}

Now instead of considering infinite space, we assume that the droplet is at the origin, in the middle of a channel of width $W$. The tangential velocity must vanish at the channelwalls. To enforce this condition, we we can use the method of images to compute the complex potential, and  the corresponding complex velocity
\begin{align}
    w(z) &= \sum_{k=-\infty}^\infty w_1(z-ikW) = \sigma \frac{\pi}{W} \coth\left(\frac{\pi z}{W}\right), \\
    v(z) &= -\sigma \left(\frac{\pi}{W}\right)^2 \sinh^{-2}\left(\frac{\pi z}{W}\right).
\end{align}
In terms of the coordinates $(x, y)$, this gives
\begin{equation}
    \vv(x, y) = \frac{-2\sigma (\pi/W)^2}{\left[\cosh(2\pi x/W)-\cos(2\pi y/W)\right]^2} \begin{pmatrix}
        \cosh(2\pi x/W) \cos(2\pi y/W) - 1 \\
        \sinh(2\pi x/W) \sin(2\pi y/W)
    \end{pmatrix}.
\end{equation}

Ignoring the fluctuations transverse to the $x$ axis, we only need to consider
\begin{equation} \label{eq:screened}
    v_x(x, 0) = -\sigma \left(\frac{\pi}{W}\right)^2 \mathrm{csch}^2\left(\frac{\pi x}{W}\right).
\end{equation}
This velocity fields acts as an effective distance-dependant force on the other droplets. This explains our choice $f_H(r) = -\left(\frac{\pi}{W}\right)^2 \mathrm{csch}^2\left(\frac{\pi r}{W}\right)$, where we set the strength to $\sigma=1$ without loss of generality.

Two comments are in order: (i) since $v_x(x, 0) = v_x(-x, 0)$, the hydrodynamic interactions are strongly non-reciprocal, this fact is at the heart of the phenomenology that we detail in the main text ; (ii) the finite width of the channel implies a screening of the interactions: the hydrodynamic forces are short-ranged (exponential decay).

\bibliographystyle{apsrev4-1}
\bibliography{referencesAlexis.bib}